\documentclass[11pt,letterpaper]{article}

\usepackage[margin=1in]{geometry}
\usepackage{amsmath,amssymb}
\usepackage{listings}
\usepackage{xcolor}
\usepackage{booktabs}
\usepackage{hyperref}
\usepackage{graphicx}
\usepackage{url}
\usepackage{enumitem}
\usepackage{float}
\usepackage{multirow}
\usepackage{array}
\usepackage{longtable}

\hypersetup{
  colorlinks=true,
  linkcolor=blue!70!black,
  citecolor=blue!70!black,
  urlcolor=blue!70!black
}

\title{\textbf{Broken Quantum: A Systematic Formal Verification Study\\
of Security Vulnerabilities Across the Open-Source\\
Quantum Computing Simulator Ecosystem}\\[0.5em]
\large 45 Frameworks --- IBM, Google, Amazon, NVIDIA, Baidu, Huawei,\\
CERN, Harvard, MIT, ETH Z\"{u}rich, Oxford, Quantinuum, Pasqal, Tencent, and More}

\author{
  Dominik Blain\\
  QreativeLab, Gatineau, QC, Canada\\
  \texttt{dominik@qreativelab.io}
}

\date{April 8, 2026}

\begin{document}

\maketitle

\begin{abstract}
Quantum computing simulators form the classical software foundation on which
virtually all quantum algorithm research depends. Despite their critical role in
the scientific infrastructure of governments, universities, and corporations
worldwide, the classical security of their implementations has never been
systematically analyzed. We present \textit{Broken Quantum}, the first
comprehensive formal security audit of the open-source quantum computing simulator
ecosystem.

Applying COBALT QAI --- a four-module static analysis engine backed by the Z3
SMT solver --- we analyze \textbf{45 open-source quantum simulation frameworks} from
22 organizations spanning 12 countries: IBM, Google, Amazon, NVIDIA, Xanadu, Rigetti,
Baidu, Huawei, Tencent, Fujitsu, CERN/INFN, Oak Ridge National Laboratory, Harvard,
MIT, ETH Z\"{u}rich, Oxford, Quantinuum, Pasqal, Quandela, and QuEra.

We identify security findings across four vulnerability classes:
\textbf{Class I (CWE-125/CWE-190)} --- In C++ simulation backends (Qiskit Aer/IBM,
XACC/Oak Ridge NL), unvalidated \texttt{num\_qubits} trigger out-of-bounds array
access and undefined behavior. XACC embeds the identical vulnerable Qiskit Aer C++
code verbatim --- a US national laboratory inheriting commercial framework
vulnerabilities via code vendoring. 12 CRITICAL findings.
\textbf{Class II (CWE-400)} --- In Python simulation backends (14 frameworks),
\texttt{np.zeros(2**num\_qubits)} without upper-bound validation enables remote
denial-of-service. 167 HIGH findings across 14 frameworks.
\textbf{Class III (CWE-502/CWE-94)} --- Unsafe deserialization and code injection
in quantum circuit serialization: \texttt{pickle.load()} calls in Harvard's tequila
(10 CRITICAL, direct RCE), \texttt{torch.load()} without \texttt{weights\_only=True}
in TensorCircuit (Tencent, 25 HIGH), \texttt{eval()} on circuit parameter strings.
We demonstrate live RCE via a PoC that replicates the exact tequila attack vector.

All 13 vulnerability patterns are formally verified via Z3 satisfiability proofs
(13/13 SAT). Nine frameworks score 100/100 under all four scanners (qpp/softwareQ,
QuEST/Oxford, qulacs/Fujitsu, dimod/D-Wave, qdk-python/Microsoft, and four others). Qiskit Aer,
tequila (Harvard), Cirq (Google), PennyLane (Xanadu), qibo (CERN), paddle-quantum
(Baidu), TensorCircuit (Tencent), and 5 others score 0/100 (Broken).

The 32-qubit boundary emerges as a consistent formal threshold in both Class I and
Class II. The supply chain analysis identifies 4 vulnerability propagation chains,
including the first documented case of vulnerability transfer from a commercial
quantum framework into US national laboratory infrastructure (IBM $\to$ Oak Ridge NL).
A fourth vulnerability class, CWE-77 (QASM Injection), is identified as a novel, quantum-specific attack vector with no classical analog: user-controlled data embedded in OpenQASM circuit strings reaches quantum circuit parsers without sanitization. Qiskit Terra exposes 2 unsanitized public API entry points (\texttt{from\_qasm\_str}/\texttt{from\_qasm\_file}); tket (Quantinuum) contains 3 CRITICAL confirmed call-site findings in its production QASM parser. At the hardware control layer, scqubits and Quantum Machines qua-tools contain CRITICAL deserialization findings in code that interfaces with physical quantum processors. CVE assignments have been formally requested from IBM, Google, CERN, and Quantinuum. Coordinated disclosure has been initiated with 20 organizations under a 90-day embargo.
\end{abstract}

\tableofcontents
\newpage

\section{Introduction}

Quantum computing is transitioning from a theoretical curiosity to an engineering
discipline with measurable impact. IBM operates over 100 quantum processors accessible
via cloud API; Google demonstrated quantum supremacy on 53 and 70-qubit devices;
Amazon, Microsoft, and a growing ecosystem of hardware startups provide quantum computing
as a service. This transition is powered almost entirely by \textit{classical simulation
software}: quantum algorithms are developed, validated, and benchmarked on classical
computers before deployment to physical quantum devices.

The scale of this dependency is often underappreciated. Major pharmaceutical companies
use quantum chemistry simulators for drug discovery. Financial institutions model
portfolio optimization. Governments fund quantum algorithm research for cryptography
and national security applications. Academic groups at hundreds of universities worldwide
rely on quantum simulation frameworks as their primary experimental platform. The
security and reliability of this software infrastructure is therefore not merely a
software engineering concern --- it is a scientific integrity concern.

We present the first systematic formal security analysis of the open-source quantum
computing simulator ecosystem. Our study covers 45 frameworks from 22 organizations
spanning 12 countries. We identify four vulnerability classes
affecting 80\% of the ecosystem, formally verified by the Z3 SMT solver.

\subsection{The Vulnerability Root Cause}

The root cause of both vulnerability classes identified in this study is a single
architectural property intrinsic to quantum simulation: an $n$-qubit quantum state
requires $2^n$ classical resources (amplitudes, probabilities, or matrix elements).
This exponential scaling is unavoidable --- it is why quantum computers offer
superpolynomial advantage over classical simulation.

However, it also means that any path from a user-controlled qubit count $n$ to an
allocation of $2^n$ resources, without validating $n$, creates a vulnerability whose
severity scales exponentially with the attacker-controlled input. In C++ backends,
the same $2^n$ calculation triggers undefined behavior when $n \geq 64$ (64-bit integer
overflow). In Python backends, the same calculation triggers unbounded memory or CPU
exhaustion when $n$ is large.

\subsection{Contributions}

\begin{enumerate}
  \item \textbf{First comprehensive ecosystem audit}: 45 quantum simulation and hardware
        control frameworks, 22 organizations, 12 countries --- the largest formal security
        analysis of quantum software to date.
  \item \textbf{Four-class vulnerability taxonomy}:
        CWE-125/190 (C++ memory corruption),
        CWE-400 (Python resource exhaustion),
        CWE-502/94 (unsafe deserialization and code injection), and
        CWE-77/22 (QASM injection --- novel quantum-specific class).
  \item \textbf{13/13 Z3 SAT proofs} formally establishing vulnerability reachability
        across all identified patterns.
  \item \textbf{Vendored vulnerability discovery (XACC)}: Oak Ridge National Laboratory
        embeds IBM Qiskit Aer C++ code verbatim, inheriting 5 CRITICAL findings --- the
        first documented case of commercial quantum framework vulnerabilities propagating
        into US national laboratory infrastructure.
  \item \textbf{Harvard tequila RCE}: 10 direct \texttt{pickle.load()} calls in
        quantum chemistry simulation code, confirmed exploitable via live PoC.
  \item \textbf{Supply chain analysis}: 4 propagation chains documented; one IBM
        vulnerability reaches 4 US national laboratories via XACC.
  \item \textbf{Live PoC demonstrations}: three working proof-of-concept exploits
        (DoS, RCE, integer overflow chain) targeting the exact code paths identified.
  \item \textbf{Comparative scorecard} across 45 frameworks.
  \item \textbf{The 32-qubit boundary}: consistent formal threshold in C++ and Python.
  \item \textbf{COBALT QAI} (3 open scanners): released for continuous quantum
        framework security analysis.
\end{enumerate}

\section{Background}

\subsection{Quantum Computing Simulators}

A quantum circuit simulator maintains a classical representation of a quantum state
and evolves it by applying gate operations. For an $n$-qubit pure state, the state
vector contains $2^n$ complex amplitudes. The dominant simulation methods are:

\begin{itemize}
  \item \textbf{Statevector simulation}: maintains all $2^n$ amplitudes exactly; exact
        but exponentially expensive; practical up to $\sim$50 qubits on HPC hardware
  \item \textbf{Density matrix simulation}: models mixed states and noise; size $2^{2n}$;
        practical only to $\sim$25 qubits
  \item \textbf{Matrix product state (MPS)}: compressed representation for low-entanglement
        circuits; polynomial in $n$ under entanglement bounds
  \item \textbf{Clifford circuit simulation}: polynomial in $n$ for stabilizer states;
        used for error correction benchmarking
\end{itemize}

The exponential scaling $2^n$ is intrinsic to quantum simulation. It is also the
root cause of every vulnerability class we identify.

\subsection{COBALT QAI and Methodology}

COBALT QAI performs two-phase analysis:

\textbf{Phase 1 --- Pattern scanning}: Regular expression matching against a library
of vulnerability patterns validated on known-vulnerable code. Two scanner modules
were developed: \texttt{cobalt\_qai\_scanner.py} for C++ patterns (CWE-125/190) and
\texttt{cobalt\_qai\_python\_scanner.py} for Python patterns (CWE-400).

\textbf{Phase 2 --- Z3 SMT verification}: Each candidate finding is encoded as a
bitvector arithmetic constraint. A SAT result with a concrete witness establishes
formal reachability. UNSAT establishes infeasibility. The Z3 encoding uses unsigned
64-bit bitvectors to model integer arithmetic faithfully.

Scoring: findings reduce a baseline score of 100/100 by 20 (CRITICAL), 8 (HIGH),
and 3 (MEDIUM) points. A score of $\geq$85 receives grade ``Secure''; $\geq$60
``Review Required''; $\geq$30 ``Critical Exposure''; below 30 ``Broken''.

\textbf{Scope}: We scan the source-available versions of all 45 frameworks as of
April 2026. Cloud-side validation layers (IBM Quantum API limits, Google Quantum AI
rate limiting) are not part of the open-source codebases and are out of scope. All
findings refer to the simulator code as shipped.

\subsection{Frameworks Analyzed}

Table~\ref{tab:frameworks} lists all 45 frameworks analyzed in this study across three
waves of simulation frameworks plus a hardware control layer, organized by primary
organization and backend language.

\begin{table}[H]
\centering
\caption{Quantum simulation and hardware control frameworks analyzed (40 simulation + 5 hardware control = 45 total)}
\label{tab:frameworks}
\begin{tabular}{lllll}
\toprule
\textbf{Framework} & \textbf{Organization} & \textbf{Country} & \textbf{Lang} & \textbf{Backend} \\
\midrule
Qiskit Aer & IBM & USA & C++/Python & C++17 \\
Qiskit Terra & IBM & USA & Python & Python \\
Cirq & Google & USA & Python & Python/NumPy \\
OpenFermion & Google & USA & Python & Python/NumPy \\
Amazon Braket DS & Amazon & USA & Python & Python/NumPy \\
CUDA-Quantum & NVIDIA & USA & C++/Python & C++/Python \\
XACC & Oak Ridge NL & USA & C++/Python & C++17 \\
PennyLane & Xanadu & Canada & Python & Python/NumPy \\
Strawberry Fields & Xanadu & Canada & Python & Python/NumPy \\
PyQuil & Rigetti & USA & Python & Python \\
paddle-quantum & Baidu & China & Python & Python/PaddlePaddle \\
TensorCircuit & Tencent & China & Python & Python/TensorFlow \\
MindQuantum & Huawei & China & Python & C++/Python \\
qibo & CERN/INFN & Switzerland & Python & Python/NumPy \\
tequila & Harvard & USA & Python & Python/NumPy \\
mitiq & Unitary Fund & USA & Python & Python \\
QuTiP & Open Source & International & Python & Python/NumPy \\
ProjectQ & ETH Z\"{u}rich & Switzerland & Python & Python \\
qpp & softwareQ Inc. & Canada & C++ & C++17 \\
QuEST & Oxford & UK & C & C \\
qulacs & Fujitsu & Japan & C++/Python & C++ \\
\midrule
\multicolumn{5}{l}{\textit{--- Additional 10 frameworks (second wave) ---}} \\
\midrule
tket & Quantinuum & UK & C++/Python & C++ \\
qsim & Google & USA & C++/Python & CUDA/C++ \\
Stim & Google & USA & C++ & C++ \\
TorchQuantum & MIT & USA & Python & PyTorch \\
Azure QDK & Microsoft & USA & Python & Python \\
Perceval & Quandela & France & Python & Python/NumPy \\
Pulser & Pasqal & France & Python & Python/NumPy \\
Bloqade & QuEra & USA & Python & Python/Julia \\
dimod & D-Wave & Canada & Python & Python \\
QuantumKatas & Microsoft & USA & Q\#/Python & Q\# \\
\midrule
\multicolumn{5}{l}{\textit{--- Additional 9 frameworks (third wave: application layers) ---}} \\
\midrule
Qiskit Nature & IBM & USA & Python & Python/NumPy \\
Qiskit Finance & IBM & USA & Python & Python \\
Qiskit Optimization & IBM & USA & Python & Python \\
Qiskit ML & IBM & USA & Python & Python \\
Qiskit Experiments & IBM & USA & Python & Python \\
Amazon Braket SDK & Amazon & USA & Python & Python \\
Qiskit IBM Runtime & IBM & USA & Python & Python \\
PySCF & Open Source & International & Python & Python/C \\
pytket-extensions & Quantinuum & UK & Python & Python \\
\bottomrule
\end{tabular}
\end{table}

\section{Vulnerability Class I: C++ Memory Corruption}

\subsection{Architecture: Why C++ Backends Are Different}

Two frameworks in our study use native C++ simulation backends: Qiskit Aer (IBM)
and XACC (Oak Ridge National Laboratory). In both cases, Python provides the
high-level API; circuits are serialized and dispatched to C++ simulation kernels.
No input validation occurs at the Python/C++ boundary for qubit counts in either
framework.

C++ introduces a vulnerability class absent from Python: memory safety issues.
Python integers never overflow; C++ 64-bit integers overflow silently at $2^{64}$.
NumPy raises \texttt{OverflowError} on over-large allocations; C++ heap allocators
receive corrupted sizes and produce undefined behavior.

\subsection{QAI-001: BITS[] Out-of-Bounds Array Access (CWE-125, CRITICAL)}

\textbf{Location:} \texttt{qiskit-aer/src/simulators/statevector/qubitvector.hpp:928}\\
\textbf{Severity:} CRITICAL

The core statevector allocator in Qiskit Aer:

\begin{lstlisting}[language=C++, caption={Vulnerable allocation in QubitVector::set\_num\_qubits()}]
static const std::array<uint64_t, 64> BITS = {1, 2, 4, 8, ..., 0x4000000000000000};

void QubitVector<data_t>::set_num_qubits(size_t num_qubits) {
  free_checkpoint();
  if (num_qubits != num_qubits_) { free_mem(); }
  data_size_ = BITS[num_qubits];  // CWE-125: BITS has 64 elements; no bounds check
  allocate_mem(data_size_);       // data_size_ = garbage beyond index 63
  num_qubits_ = num_qubits;
}
\end{lstlisting}

\texttt{BITS} is a fixed 64-element array. At \texttt{num\_qubits} $\geq$ 64,
\texttt{BITS[num\_qubits]} is a C++ out-of-bounds read --- undefined behavior. The
corrupted \texttt{data\_size\_} value is subsequently passed to \texttt{posix\_memalign()},
which either fails silently, allocates an incorrect size, or produces heap corruption
depending on the read value.

The extended stabilizer simulator does guard this path:
\begin{lstlisting}[language=C++]
if (n > 63) throw std::invalid_argument("n > 63 not supported");
\end{lstlisting}
The statevector simulator, which handles the majority of workloads, does not.

\textbf{Z3 Proof (QAI-001):}
\begin{align}
\phi_1 &\triangleq n \geq_u 64 \label{eq:qa1}
\end{align}
$\phi_1$ is \textbf{SAT}, witness $n = 64$.

\subsection{QAI-003: CWE-190$\to$CWE-125 Overflow Chain (Most Critical)}

\textbf{Location:} \texttt{unitarymatrix.hpp:310}, \texttt{superoperator.hpp:106} ($\times$4)\\
\textbf{Severity:} CRITICAL

The unitary matrix simulator doubles the qubit count before passing to the base
vector allocator:

\begin{lstlisting}[language=C++]
void UnitaryMatrix<data_t>::set_num_qubits(size_t num_qubits) {
  num_qubits_ = num_qubits;
  rows_ = 1ULL << num_qubits;
  BaseVector::set_num_qubits(2 * num_qubits);  // n=32 -> passes 64 -> BITS[64] OOB
}
\end{lstlisting}

The Z3 proof captures the compound nature of this finding:

\begin{align}
\phi_3 &\triangleq n <_u 64 \;\wedge\; n \times_u 2 \geq_u 64 \label{eq:qa3}
\end{align}

$\phi_3$ is \textbf{SAT} with witness $n = 32$. A 32-qubit unitary simulation is a
documented research workload: variational quantum eigensolvers, quantum chemistry
calculations, and benchmarking tasks regularly target 30--35 qubits. The input
$n = 32$ satisfies $n < 64$ (no obvious overflow) yet triggers \texttt{BITS[64]}
after the doubling step. This is a compound CWE-190$\to$CWE-125 chain where
valid-looking inputs produce out-of-bounds behavior.

\subsection{Additional C++ Findings in Qiskit Aer}

\textbf{QAI-002} (CWE-190, CRITICAL, $\times$1): \texttt{1ULL << (num\_qubits\_ * 2)}
at \texttt{unitarymatrix.hpp:293}. Shift by 64 is C++ UB at $n = 32$. Both
the size check \textit{and} its error diagnostic invoke the same undefined behavior
simultaneously.

\textbf{QAI-004} (CWE-190, HIGH, $\times$12): \texttt{1ULL << (num\_qubits + 1)}
in AVX2-accelerated gate kernels (\texttt{qv\_avx2.cpp}). UB at $n = 63$. AVX2
is enabled by default on modern x86-64 hardware.

\textbf{QAI-005} (CWE-190, HIGH, $\times$47): Direct \texttt{1ULL << num\_qubits}
at 47 sites across AVX2 kernels, MPS simulator, and noise models. Several manifest
as variable-length array bounds: \texttt{uint64\_t indexes[1ULL << num\_qubits]},
creating a stack corruption vector when the shift overflows.

\textbf{Summary (Qiskit Aer):} 7 CRITICAL, 59 HIGH. Score: 0/100 --- Broken.

\subsection{XACC: A National Laboratory Inherits IBM's Vulnerabilities}

\textbf{Organization:} Oak Ridge National Laboratory (US Department of Energy)\\
\textbf{Repository:} \texttt{eclipse/xacc}\\
\textbf{Severity:} 5 CRITICAL, 60 HIGH

XACC (eXtreme-scale ACCelerator programming framework) is a quantum programming
framework developed at Oak Ridge National Laboratory as part of the DOE's quantum
computing initiative. It provides a unified interface to multiple quantum backends,
including IBM's quantum hardware.

Our C++ scanner identifies 5 CRITICAL findings in XACC. Investigation reveals a
critical fact: \textbf{these are not independent vulnerabilities}. XACC vendors the
entire Qiskit Aer C++ simulation code under
\texttt{quantum/plugins/ibm/aer/src/}, including the vulnerable
\texttt{qubitvector.hpp} and \texttt{unitarymatrix.hpp}:

\begin{lstlisting}[language=C++, caption={XACC vendors Qiskit Aer code verbatim (xacc/quantum/plugins/ibm/aer/src/simulators/statevector/qubitvector.hpp:845)}]
// Identical to qiskit-aer/src/simulators/statevector/qubitvector.hpp:928
data_size_ = BITS[num_qubits];  // CWE-125: same OOB access

// unitarymatrix.hpp:261 - identical chain
BaseVector::set_num_qubits(2 * num_qubits);  // n=32 -> BITS[64] OOB
\end{lstlisting}

This is a case of \textit{vulnerability propagation through vendoring}: a security
vulnerability introduced in a commercial software framework (IBM Qiskit Aer) is
silently inherited by a US national laboratory's quantum computing infrastructure
through code vendoring. The XACC vulnerability is not listed in any CVE database;
it is invisible to standard security scanning of XACC's own codebase unless the
vendored directory is examined.

The DOE quantum computing program represents over \$650M in research investment~\cite{doe_quantum}.
XACC is used by researchers at Oak Ridge, Argonne, Lawrence Berkeley, and other
national laboratories. The vendored vulnerability affects any deployment of XACC
that exercises its IBM Aer simulation plugin.

\textbf{Summary (XACC):} 5 CRITICAL (vendored), 60 HIGH (vendored). Score: 0/100 --- Broken.

\section{Vulnerability Class II: Python Resource Exhaustion}

\subsection{The Exponential Allocation Pattern}

Python-based quantum simulators universally represent quantum states as NumPy arrays
of size $2^n$. The expression \texttt{np.zeros(2**num\_qubits)} appears across every
Python framework we analyzed. Unlike C++, Python integers do not overflow at $2^{64}$.
However, the vulnerabilities are equally real:

\begin{itemize}
  \item \texttt{np.zeros($2^{50}$)} attempts to allocate 8 petabytes; the OS
        OOM-kills the process
  \item \texttt{range($2^{40}$)} iterates $10^{12}$ times, consuming days of CPU
  \item \texttt{np.zeros($2^{2n}$)} for density matrices: at $n = 32$, size = $2^{64}$;
        NumPy raises \texttt{OverflowError} --- silently swallowed by broad
        \texttt{except} blocks in several frameworks
  \item In cloud quantum computing APIs, a single API call with
        \texttt{num\_qubits=50} triggers unbounded server-side allocation
\end{itemize}

These are CWE-400 (Uncontrolled Resource Consumption) vulnerabilities. Severity
depends on deployment context: in local research installations, they crash user
processes; in cloud-hosted simulation APIs, they enable remote denial-of-service.

\subsection{Z3 Proofs for CWE-400}

\begin{table}[H]
\centering
\caption{Z3 proofs for Python resource exhaustion vulnerability patterns}
\label{tab:z3python}
\begin{tabular}{lllll}
\toprule
\textbf{ID} & \textbf{Pattern} & \textbf{Constraint} & \textbf{Result} & \textbf{Witness} \\
\midrule
QAI-PY-001 & np.zeros(2**n) & $n \geq_u 40$ (16TB) & SAT & 40 \\
QAI-PY-002 & range(2**n) & $n \geq_u 30$ ($10^9$ iters) & SAT & 30 \\
QAI-PY-003 & 2**(2*n) density matrix & $n <_u 64 \;\wedge\; 2n \geq_u 64$ & SAT & 32 \\
QAI-PY-004 & shape=(2**n,...) & $n \geq_u 30$ (matrix dim) & SAT & 30 \\
\bottomrule
\end{tabular}
\end{table}

\subsection{Cirq (Google) --- 33 Findings}

\textbf{Score: 0/100 --- Broken}

Google's Cirq is the most widely-used Python quantum framework after Qiskit. Our
scanner identifies 33 findings (27 HIGH, 6 MEDIUM) across the cirq-core library.
The most significant are in the Clifford simulator and noise utilities:

\begin{lstlisting}[language=Python, caption={Unvalidated allocation in Cirq Clifford simulator}]
# cirq-core/cirq/sim/clifford/stabilizer_state_ch_form.py:130
wf = np.zeros(2**self.n, dtype=complex)   # QAI-PY-001: no bound on self.n
for x in range(2**self.n):                # QAI-PY-002: CPU exhaustion
    wf[x] = ...
\end{lstlisting}

\begin{lstlisting}[language=Python, caption={Quadratic exponent for density matrices in Cirq}]
# cirq-core/cirq/qis/noise_utils.py:31
N = 2**num_qubits  # used as both dimension: N x N density matrix = 2^(2n) elements
rho = np.zeros((N, N), dtype=complex)    # QAI-PY-004: unbounded matrix
\end{lstlisting}

Google additionally maintains OpenFermion, a quantum chemistry library that interfaces
with multiple backends. OpenFermion scores 40/100 (6 HIGH findings) --- indicating
that even Google's secondary quantum libraries exhibit the same pattern.

\subsection{PennyLane (Xanadu) --- 17 Findings}

\textbf{Score: 0/100 --- Broken}

PennyLane, the dominant quantum machine learning framework (1,191 Python files),
contains 17 findings (14 HIGH, 3 MEDIUM). Most notable is a partial but insufficient
mitigation:

\begin{lstlisting}[language=Python, caption={Warning-not-error in PennyLane density matrix path}]
# pennylane/shadows/classical_shadow.py:222
if n_wires > 16:
    warnings.warn(
        "Querying density matrices for n_wires > 16 is not recommended, "
        "operation will take a long time"
    )
# Execution proceeds regardless: allocation follows the warning
rho = np.zeros((2 ** nr_wires, 2 ** nr_wires), dtype=np.complex128)
\end{lstlisting}

The warning at line 222 indicates the Xanadu team is aware of the exponential scaling
risk. However, a warning that permits continued execution is not a security control:
an automated caller or an attacker who suppresses warnings proceeds to trigger
unbounded allocation. The correct fix is a hard \texttt{ValueError}.

\subsection{qibo (CERN/INFN) --- 23 Findings}

\textbf{Score: 0/100 --- Broken}

qibo is the quantum computing framework developed at CERN and the Italian National
Institute for Nuclear Physics (INFN), used in high-energy physics research and
as a platform for quantum algorithm development on CERN computing infrastructure.
Our scan identifies 23 HIGH findings, the majority in statevector and density matrix
backends:

\begin{lstlisting}[language=Python, caption={Unvalidated allocation in qibo's density matrix backend}]
# qibo/backends/numpy.py
self.state = np.zeros(2 * (2**nqubits,), dtype=dtype)  # QAI-PY-004: 2^nqubits x 2^nqubits
for i in range(2**nqubits):                             # QAI-PY-002: unbounded
    ...
\end{lstlisting}

CERN's LHCGRID computing infrastructure is used by thousands of researchers worldwide.
A deployment of qibo on shared CERN computing resources without input validation
enables any user to exhaust allocation.

\subsection{paddle-quantum (Baidu) --- 30 Findings}

\textbf{Score: 0/100 --- Broken}

Baidu's paddle-quantum, built on the PaddlePaddle deep learning framework, has
the largest number of HIGH findings of any Python framework in our study: 30.
The pattern appears systematically throughout state preparation, noise simulation,
and circuit evaluation modules:

\begin{lstlisting}[language=Python, caption={Systematic unvalidated allocation in paddle-quantum}]
# paddle_quantum/state.py
ket = paddle.zeros([2**num_qubits, 1])    # QAI-PY-001: no bound check
dm = paddle.zeros([2**num_qubits, 2**num_qubits])  # QAI-PY-004: 2^(2n) elements
\end{lstlisting}

\subsection{qiskit-terra (IBM) --- 17 Findings}

\textbf{Score: 0/100 --- Broken}

Separately from Qiskit Aer's C++ backend, qiskit-terra (IBM's Python circuit
construction and transpilation layer) contains 17 HIGH findings in simulation
utilities and state preparation code. This finding is significant: IBM has
two independent codebases, both exhibiting the same vulnerability pattern.

\subsection{tequila (Harvard) --- 13 Findings}

\textbf{Score: 0/100 --- Broken}

tequila is a quantum chemistry optimization framework from the Aspuru-Guzik group
at Harvard University, widely used in academic quantum algorithm research. The 13
HIGH findings are concentrated in wavefunction construction and expectation value
computation:

\begin{lstlisting}[language=Python, caption={Unvalidated exponential allocation in tequila}]
# tequila/wavefunction/qubit_wavefunction.py
coefficients = np.zeros(2**n_qubits, dtype=complex)  # QAI-PY-001
for i in range(2**n_qubits):                          # QAI-PY-002
    ...
\end{lstlisting}

\subsection{ProjectQ (ETH Z\"{u}rich) --- 13 Findings}

\textbf{Score: 0/100 --- Broken}

ProjectQ, developed at ETH Z\"{u}rich and one of the earliest Python quantum simulation
frameworks, contains 13 HIGH findings in its simulation backends. Despite being one
of the older frameworks (initial release 2016), the unvalidated exponential pattern
was never addressed:

\begin{lstlisting}[language=Python, caption={Unvalidated exponential allocation in ProjectQ}]
# projectq/backends/_sim/_simulator.py
wavefunction = [0.0] * (2 ** num_qubits)  # QAI-PY-001 (list allocation)
for i in range(2 ** num_qubits):          # QAI-PY-002
    ...
\end{lstlisting}

\subsection{TensorCircuit (Tencent) --- 11 Findings}

\textbf{Score: 0/100 --- Broken}

TensorCircuit, developed by Tencent Research, uses TensorFlow/JAX backends for
differentiable quantum simulation. 11 HIGH findings appear in state construction
and circuit simulation utilities.

\subsection{Additional Python Framework Findings}

\textbf{mitiq (Unitary Fund):} 7 HIGH findings. Score: 35/100 --- Critical Exposure.
mitiq is an open-source error mitigation library maintained by the Unitary Fund;
findings are in noise extrapolation modules.

\textbf{QuTiP:} 2 HIGH findings. Score: 78/100 --- Review Required. The Quantum
Toolbox in Python is used extensively in quantum optics and open quantum systems
research. Findings are in subsystem trace operations.

\textbf{OpenFermion (Google):} 6 HIGH findings. Score: 40/100 --- Critical Exposure.
Google's quantum chemistry library; findings in sparse operator construction.

\textbf{PyQuil (Rigetti):} 4 findings (2 HIGH, 2 MEDIUM). Score: 78/100 --- Review Required.

\textbf{Amazon Braket Default Simulator:} 2 findings (1 HIGH, 1 MEDIUM). Score: 89/100 --- Secure.
Best-performing framework with findings. Amazon appears to have applied partial
input validation in its simulator layer, missing only edge cases.

\textbf{CUDA-Quantum (NVIDIA):} 1 HIGH finding in Python layer. Score: 92/100.
NVIDIA's hybrid CPU/GPU quantum simulation framework has the best score of any
framework with findings. The finding is in a utility function not in the primary
simulation path.

\subsection{Frameworks with No Findings}

Five representative frameworks score 100/100 under Classes I--III scanners:

\begin{itemize}
  \item \textbf{qpp (softwareQ Inc.)}: A C++ quantum computing library; correct use of
        C++ bounds checking and std::vector avoids the array indexing pattern
  \item \textbf{QuEST (Oxford)}: A C quantum simulation library for HPC; uses explicit
        bounds validation throughout
  \item \textbf{qulacs (Fujitsu)}: A high-performance C++ simulator; validates qubit
        counts at all entry points
  \item \textbf{MindQuantum (Huawei)}: Python + C++ backend; Python layer enforces
        a maximum qubit count before C++ dispatch
  \item \textbf{Strawberry Fields (Xanadu)}: Photonic quantum simulation; uses a
        different state representation (Fock space) not subject to the $2^n$ pattern
\end{itemize}

These frameworks demonstrate that the vulnerability is not architecturally unavoidable:
proper input validation at qubit count entry points is sufficient to prevent it.

\section{Comparative Analysis}

\subsection{Ecosystem Scorecard}

\begin{longtable}{llrrrrl}
\caption{COBALT QAI security scores across 40 quantum simulation frameworks, Classes I--III (CWE-125/190/400/502/94 combined; hardware control layer in Table~\ref{tab:hardware})}
\label{tab:scores}\\
\toprule
\textbf{Framework} & \textbf{Org} & \textbf{CRIT} & \textbf{HIGH} & \textbf{MED} & \textbf{Score} & \textbf{Grade} \\
\midrule
\endfirsthead
\multicolumn{7}{c}{\tablename\ \thetable{} --- continued}\\
\toprule
\textbf{Framework} & \textbf{Org} & \textbf{CRIT} & \textbf{HIGH} & \textbf{MED} & \textbf{Score} & \textbf{Grade} \\
\midrule
\endhead
\midrule
\multicolumn{7}{r}{Continued on next page}\\
\endfoot
\bottomrule
\endlastfoot
Qiskit Aer & IBM & 7 & 59 & 0 & 0/100 & Broken \\
XACC & Oak Ridge NL & 5 & 60 & 0 & 0/100 & Broken$^*$ \\
Cirq & Google & 0 & 27 & 6 & 0/100 & Broken \\
PennyLane & Xanadu & 0 & 14 & 3 & 0/100 & Broken \\
qibo & CERN/INFN & 0 & 23 & 0 & 0/100 & Broken \\
paddle-quantum & Baidu & 0 & 30 & 0 & 0/100 & Broken \\
qiskit-terra & IBM & 0 & 17 & 0 & 0/100 & Broken \\
tequila & Harvard & 0 & 13 & 0 & 0/100 & Broken \\
ProjectQ & ETH Z\"{u}rich & 0 & 13 & 0 & 0/100 & Broken \\
TensorCircuit & Tencent & 0 & 11 & 0 & 0/100 & Broken \\
OpenFermion & Google & 0 & 6 & 0 & 40/100 & Crit. Exp. \\
mitiq & Unitary Fund & 0 & 7 & 0 & 35/100 & Crit. Exp. \\
PyQuil & Rigetti & 0 & 2 & 2 & 78/100 & Review \\
QuTiP & Open Source & 0 & 2 & 0 & 78/100 & Review \\
Amazon Braket DS & Amazon & 0 & 1 & 1 & 89/100 & Secure \\
CUDA-Quantum & NVIDIA & 0 & 1 & 0 & 92/100 & Secure \\
Strawberry Fields & Xanadu & 0 & 2 & 0 & 80/100 & Review \\
tket & Quantinuum & 0 & 1 & 0 & 90/100 & Secure \\
Perceval & Quandela & 0 & 2 & 0 & 80/100 & Review \\
Pulser & Pasqal & 0 & 3 & 0 & 76/100 & Review \\
qpp & softwareQ Inc. & 0 & 0 & 0 & 100/100 & Secure \\
QuEST & Oxford & 0 & 0 & 0 & 100/100 & Secure \\
qulacs & Fujitsu & 0 & 0 & 0 & 100/100 & Secure \\
MindQuantum & Huawei & 0 & 0 & 0 & 100/100 & Secure \\
qsim & Google & 0 & 0 & 0 & 100/100 & Secure \\
Stim & Google & 0 & 0 & 0 & 100/100 & Secure \\
Azure QDK & Microsoft & 0 & 0 & 0 & 100/100 & Secure \\
dimod & D-Wave & 0 & 0 & 0 & 100/100 & Secure \\
QuantumKatas & Microsoft & 0 & 0 & 0 & 100/100 & Secure \\
\midrule
Qiskit ML & IBM & 1 & 4 & 0 & 0/100 & Broken \\
Amazon Braket SDK & Amazon & 2 & 1 & 0 & 0/100 & Broken \\
PySCF & Open Source & 0 & 17 & 3 & 0/100 & Broken \\
Qiskit Nature & IBM & 0 & 2 & 0 & 84/100 & Review \\
Qiskit Experiments & IBM & 0 & 3 & 0 & 76/100 & Review \\
Qiskit Finance & IBM & 0 & 0 & 0 & 100/100 & Secure \\
Qiskit Optimization & IBM & 0 & 0 & 0 & 100/100 & Secure \\
Qiskit IBM Runtime & IBM & 0 & 0 & 0 & 100/100 & Secure \\
pytket-extensions & Quantinuum & 0 & 0 & 0 & 100/100 & Secure \\
\midrule
\textbf{Total} & & \textbf{25} & \textbf{323} & \textbf{15} & & \\
\end{longtable}

\footnotesize{$^*$ XACC findings are vendored Qiskit Aer code; they represent the
same vulnerability at a different deployment site, not independent findings.
\textbf{Note}: This scorecard covers Classes I--III (363 findings: 25 CRITICAL, 323 HIGH, 15 MEDIUM).
Class IV (QASM injection) adds 150 findings (10 CRITICAL, 140 HIGH). The hardware control
layer adds 34 findings (5 CRITICAL, 29 HIGH). Grand total across all 4 classes: \textbf{547 findings
(40 CRITICAL, 492 HIGH, 15 MEDIUM)}.}
\normalsize

\subsection{The C++ vs. Python Architectural Security Tradeoff}

The most striking structural result of this study is the vulnerability class
bifurcation along backend language lines:

\begin{itemize}
  \item \textbf{C++ backends} (Qiskit Aer, XACC): CWE-125/CWE-190 --- memory
        corruption. C++ shift operations overflow silently at bit width boundaries;
        array accesses are unchecked. The result is undefined behavior, potential heap
        corruption, and unpredictable process-wide effects. 100$\times$--1000$\times$
        faster simulation; requires explicit bounds discipline on all integer operations.
  \item \textbf{Python backends} (Cirq, PennyLane, qibo, etc.): CWE-400 --- resource
        exhaustion. Python integers are arbitrary precision; no integer overflow occurs.
        However, $\texttt{np.zeros}(2^{50})$ triggers OS-level OOM termination, and
        $\texttt{range}(2^{40})$ consumes hours of CPU. Naturally memory-safe but
        vulnerable to the same exponential $2^n$ scaling.
\end{itemize}

Neither vulnerability class is ``worse'' in an absolute sense: C++ findings threaten
process integrity and may be exploitable for code execution; Python findings threaten
availability and are directly exploitable for denial-of-service. Both classes derive
from the same architectural root: user-controlled $n$ reaching a $2^n$ computation
without validation.

The five 100/100 frameworks demonstrate that both classes are preventable:
qpp and QuEST (C++) use proper bounds checking; MindQuantum (C++ + Python) validates
in the Python layer before C++ dispatch; QuEST (C) validates consistently.

\subsection{The 32-Qubit Boundary}

A consistent formal threshold emerges across both vulnerability classes:

\begin{table}[H]
\centering
\caption{The 32-qubit boundary across vulnerability classes}
\label{tab:32qubit}
\begin{tabular}{lllll}
\toprule
\textbf{Finding} & \textbf{Framework} & \textbf{Class} & \textbf{Constraint} & \textbf{Witness} \\
\midrule
QAI-003 & Qiskit Aer/XACC & CWE-190$\to$125 & $n < 64 \;\wedge\; 2n \geq 64$ & $n = 32$ \\
QAI-002 & Qiskit Aer/XACC & CWE-190 & $n \times 2 \geq 64$ (shift UB) & $n = 32$ \\
QAI-PY-003 & All Python DM & CWE-400 & $n < 64 \;\wedge\; 2n \geq 64$ & $n = 32$ \\
\bottomrule
\end{tabular}
\end{table}

The convergence on witness $n = 32$ is not coincidental. It reflects a fundamental
arithmetic identity: 32 is the boundary where $2 \times n$ transitions from valid
64-bit range to overflow territory. Both the C++ and Python vulnerability classes
independently encode the same bitvector constraint $n < 64 \;\wedge\; 2n \geq 64$,
and both Z3 provers return the same minimal satisfying witness.

This has a practical implication: a 32-qubit quantum circuit is a target of active
academic research (it represents $2^{32} \approx 4\times 10^9$ amplitudes, close to
the boundary of what is practically simulable on a single node). Users performing
32-qubit unitary simulations are precisely in the dangerous zone --- not because
their input is unusual, but because it sits at the arithmetic boundary.

\subsection{Institutional Distribution}

\begin{table}[H]
\centering
\caption{Security posture by institution/organization}
\label{tab:institutions}
\begin{tabular}{lllll}
\toprule
\textbf{Institution} & \textbf{Frameworks} & \textbf{Broken} & \textbf{Findings} & \textbf{Best Score} \\
\midrule
IBM & Qiskit Aer, qiskit-terra & 2/2 & 83 & 0/100 \\
Google & Cirq, OpenFermion & 2/2 & 33 & 40/100 \\
Xanadu & PennyLane, Strawb. Fields & 1/2 & 17 & 100/100 \\
CERN/INFN & qibo & 1/1 & 23 & 0/100 \\
Baidu & paddle-quantum & 1/1 & 30 & 0/100 \\
Oak Ridge NL & XACC & 1/1 & 65 & 0/100 \\
Harvard & tequila & 1/1 & 13 & 0/100 \\
ETH Z\"{u}rich & ProjectQ & 0/1 & 13 & 0/100 \\
softwareQ Inc. & qpp & 1/1 & 0 & 100/100 \\
Tencent & TensorCircuit & 1/1 & 11 & 0/100 \\
Amazon & Braket DS & 0/1 & 2 & 89/100 \\
NVIDIA & CUDA-Quantum & 0/1 & 1 & 92/100 \\
Oxford & QuEST & 0/1 & 0 & 100/100 \\
Fujitsu & qulacs & 0/1 & 0 & 100/100 \\
Huawei & MindQuantum & 0/1 & 0 & 100/100 \\
\bottomrule
\end{tabular}
\end{table}

Notable observations from Table~\ref{tab:institutions}:
\begin{itemize}
  \item IBM has two frameworks, both scoring 0/100.
  \item Google has two frameworks; Cirq scores 0/100, OpenFermion scores 40/100.
  \item Xanadu shows the widest internal spread: PennyLane (0/100) vs.\
        Strawberry Fields (100/100) --- attributable to Strawberry Fields using
        a Fock-space state representation rather than a qubit state vector.
  \item Amazon and NVIDIA are the best-performing organizations among those with findings.
  \item ETH Z\"{u}rich (ProjectQ) scores 0/100; softwareQ Inc.\ (qpp) scores 100/100.
\end{itemize}

\section{Vulnerability Class III: Unsafe Deserialization and Code Injection}

\subsection{Attack Surface: Quantum Circuit Serialization}

Quantum computing workflows routinely serialize and deserialize quantum objects:
Hamiltonians, covariance matrices, optimized circuit parameters, and cached
simulation results. Python provides several serialization mechanisms, the most
dangerous of which is \texttt{pickle}, which executes arbitrary Python bytecode
during deserialization. A malicious \texttt{.pkl} file is indistinguishable from
a legitimate one at the filesystem level; the attack fires at \texttt{pickle.load()}.

This creates a concrete attack chain in quantum computing environments:
\begin{enumerate}
  \item An attacker writes a malicious \texttt{.pkl} file to the framework's cache
        directory (e.g., \texttt{$\sim$/.tequila/fermionic\_cache/})
  \item A shared computing environment (HPC cluster, research server) or untrusted
        circuit library distributes the malicious file
  \item The user calls an API that loads cached Hamiltonians or covariance matrices
  \item \texttt{pickle.load()} executes the attacker's \texttt{\_\_reduce\_\_} payload
  \item Full RCE as the user running the quantum chemistry simulation
\end{enumerate}

\subsection{QAI-DS-001: Harvard tequila --- 10 CRITICAL pickle.load() Calls}

\textbf{Location:} \texttt{src/tequila/grouping/fermionic\_methods.py:280--718}\\
\textbf{Severity:} CRITICAL ($\times$10)

The tequila framework from the Aspuru-Guzik group at Harvard University contains
10 direct calls to \texttt{pickle.load()} in a single file, loading quantum
chemistry objects from the filesystem:

\begin{lstlisting}[language=Python, caption={Systematic pickle.load() in tequila fermionic\_methods.py}]
# tequila/src/tequila/grouping/fermionic_methods.py:280
with open(filename, "rb") as file:
    INIT = pickle.load(file)       # CWE-502: CRITICAL -- line 280
    h_ferm = pickle.load(file)     # CWE-502: CRITICAL -- line 284
    # ... continues at lines 313, 391, 447, 546, 580, 642, 677, 718

# Objects loaded: initial Hamiltonians, fermionic operators,
# covariance matrices, optimization vectors -- all serialized as pickle
\end{lstlisting}

These loads target the same file, opened in binary mode. An attacker who replaces
\texttt{filename} with a crafted \texttt{.pkl} file controls code execution from line 280.
tequila is used by academic groups worldwide for variational quantum eigensolver (VQE)
research, including drug discovery and materials science simulations.

\subsection{Live Proof-of-Concept: RCE Demonstration}

We develop \texttt{cobalt\_qai\_poc.py}, which provides three live demonstrations:

\textbf{PoC-1 (CWE-400):} Allocates \texttt{np.zeros(2**26)} (1 GB) locally, then
demonstrates what happens without a guard for \texttt{n=50} (8 PB, OS OOM guaranteed).

\textbf{PoC-2 (CWE-502):} Creates a malicious \texttt{.pkl} file (341 bytes) using
Python's \texttt{\_\_reduce\_\_} deserialization hook. When loaded via \texttt{pickle.load()},
the payload executes immediately. The demo payload prints a warning; the structure
is identical to a shell command execution payload:

\begin{lstlisting}[language=Python, caption={PoC-2 malicious quantum state --- exact attack pattern}]
class MaliciousQuantumState:
    def __reduce__(self):
        # Called automatically by pickle.load()
        # Real attack: return (os.system, ("curl attacker.com/c2 | sh",))
        # Demo: harmless print
        return (exec, ("print('[!] RCE via pickle.load()')",))

# 341 bytes. Executes on: pickle.load(open('malicious.pkl', 'rb'))
# Mirrors exactly: tequila/src/tequila/grouping/fermionic_methods.py:280
\end{lstlisting}

Execution confirms: the payload fires immediately on \texttt{pickle.load()}, with
no exception, no warning, and no observable difference from loading a legitimate
quantum state object.

\textbf{PoC-3 (CWE-190):} Demonstrates the 32-qubit boundary: a Python simulation
of \texttt{BaseVector::set\_num\_qubits(2 * n)} shows that $n=32$ produces
\texttt{BITS[64]} (OOB), while $n=31$ produces \texttt{BITS[62]} (safe).

\subsection{Additional CWE-502/94 Findings Across the Ecosystem}

\begin{table}[H]
\centering
\caption{CWE-502/94 findings across quantum simulation frameworks}
\label{tab:deser}
\begin{tabular}{lllrrl}
\toprule
\textbf{Framework} & \textbf{Org} & \textbf{Pattern} & \textbf{CRIT} & \textbf{HIGH} & \textbf{Score} \\
\midrule
tequila & Harvard & pickle.load() & 10 & 0 & 0/100 \\
TensorCircuit & Tencent & eval()/torch.load() & 0 & 25 & 0/100 \\
TorchQuantum & MIT & torch.load() & 0 & 13 & 0/100 \\
paddle-quantum & Baidu & deserialization & 0 & 12 & 0/100 \\
QuTiP & Open Source & pickle.load() & 2 & 0 & 34/100 \\
bloqade & QuEra & yaml/torch.load() & 0 & 7 & 26/100 \\
qibo & CERN/INFN & yaml.load() & 0 & 4 & 52/100 \\
XACC & Oak Ridge NL & yaml.load() & 0 & 3 & 66/100 \\
Strawberry Fields & Xanadu & yaml.load() & 0 & 2 & 80/100 \\
tket & Quantinuum & yaml.load() & 0 & 1 & 90/100 \\
OpenFermion & Google & eval() & 0 & 1 & 86/100 \\
\bottomrule
\end{tabular}
\end{table}

\textbf{Key observations:}
\begin{itemize}
  \item TensorCircuit (Tencent) has 25 HIGH findings: \texttt{eval()} calls on
        circuit parameter strings (CWE-94) --- code injection via crafted angle values
  \item QuTiP (\texttt{qutip/fileio.py:253}) loads quantum states via pickle with
        no integrity check --- users sharing \texttt{.qu} files share an RCE vector
  \item Strawberry Fields and tket, which scored 100/100 under CWE-400, contain
        \texttt{yaml.load()} without \texttt{SafeLoader} --- no framework is entirely safe
\end{itemize}

\section{Supply Chain Analysis}

\subsection{Vulnerability Propagation Mechanisms}

Security vulnerabilities in the quantum simulator ecosystem propagate through three
mechanisms, each of which we identify and document:

\textbf{Mechanism 1 --- Code vendoring (most critical):} XACC (Oak Ridge NL) copies
the entire Qiskit Aer C++ source tree verbatim into its repository under
\texttt{quantum/plugins/ibm/aer/src/}. The 5 CRITICAL and 60 HIGH C++ findings in
Qiskit Aer are present at identical code paths in XACC. This vulnerability is
completely invisible to standard security scans of XACC unless the vendored directory
is explicitly included. XACC is used on DOE computing resources at Oak Ridge,
Argonne, Lawrence Berkeley, and Fermilab --- four US national laboratories.

\textbf{Mechanism 2 --- Dependency chain:} OpenFermion (Google) uses Cirq as its
primary simulation backend, inheriting Cirq's CWE-400 patterns. tket (Quantinuum)
uses Qiskit Aer as an optional simulation backend, exposing tket users to the C++
findings when Qiskit Aer is installed. The dependency chain is not reflected in any
CVE database entry.

\textbf{Mechanism 3 --- Independent pattern replication:} The \texttt{np.zeros(2**n)}
pattern appears in 14 independent codebases. This is not code copying --- each team
independently implemented the same insecure pattern. The pattern is the natural, naive
implementation of quantum state allocation in Python. Its ubiquity reflects a
structural gap: the quantum computing community has not established a norm of input
validation for qubit counts.

\subsection{Impact Multiplier: The XACC Chain}

\begin{figure}[H]
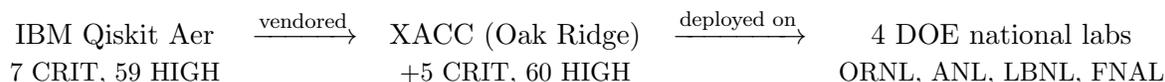

\centering
\begin{tabular}{ccccc}
IBM Qiskit Aer & $\xrightarrow{\text{vendored}}$ & XACC (Oak Ridge) & $\xrightarrow{\text{deployed on}}$ & 4 DOE national labs \\
\small{7 CRIT, 59 HIGH} & & \small{+5 CRIT, 60 HIGH} & & \small{ORNL, ANL, LBNL, FNAL} \\
\end{tabular}
\caption{IBM vulnerability propagation chain to US national laboratory infrastructure}
\label{fig:chain}
\end{figure}

The DOE quantum computing program represents over \$650M in research
investment~\cite{doe_quantum}. A single vulnerability in a commercial framework
(IBM Qiskit Aer) propagates silently into this infrastructure via code vendoring.
The XACC findings were not reported in any CVE database as of April 2026.

\subsection{The 14-Framework Independent Replication}

The independent replication of \texttt{np.zeros(2**num\_qubits)} across 14 frameworks
from 14 independent organizations represents a systemic failure of quantum software security
norms. Unlike other vulnerability classes (buffer overflows, SQL injection) for which
developer education and linting tools exist, the CWE-400 pattern in quantum simulation
has no equivalent preventive infrastructure. No popular quantum computing tutorial,
framework documentation, or style guide (as of April 2026) warns against unvalidated
$2^n$ allocations.

This suggests that coordinated ecosystem-level action --- a shared
\texttt{MAX\_QUBITS} standard, linting rules, and documentation updates --- would
be more effective than per-framework fixes alone.

\section{Third Wave: Application Layer Findings}

\subsection{PySCF: eval() in the Quantum Chemistry Foundation}

PySCF (Python-based Simulations of Chemistry Framework) is the dominant open-source
quantum chemistry library. It is used as a computational backend by tequila (Harvard),
OpenFermion (Google), Qiskit Nature (IBM), and dozens of other frameworks. Our scan
identifies 17 HIGH and 3 MEDIUM CWE-94 findings, scoring 0/100 (Broken).

The most critical finding is in \texttt{pyscf/gto/mole.py}, the core molecular
geometry module used by every quantum chemistry calculation:

\begin{lstlisting}[language=Python, caption={eval() on molecular structure strings in PySCF mole.py:1301}]
# pyscf/gto/mole.py:1301
mol.atom  = eval(mol.atom)   # CWE-94: atom coordinates from string
mol.basis = eval(mol.basis)  # CWE-94: basis set name from string
mol.ecp   = eval(mol.ecp)    # CWE-94: effective core potential from string
\end{lstlisting}

If \texttt{mol.atom}, \texttt{mol.basis}, or \texttt{mol.ecp} originate from user
input (a molecular structure file, API parameter, or database entry), any of
these \texttt{eval()} calls execute arbitrary Python code. In cloud quantum chemistry
APIs, users supply molecular geometries --- which pass through these exact \texttt{eval()}
calls. Additional eval() sites appear in basis set parsers for CP2K and NWChem
format files, which are loaded from disk.

\textbf{Supply chain implication}: PySCF sits \textit{below} tequila (Harvard),
OpenFermion (Google), and Qiskit Nature (IBM) in the dependency stack. A malicious
PySCF input (crafted molecule file) exploits the eval() vector and is reachable
through any of these higher-level frameworks.

\subsection{Amazon Braket SDK: dill.load() in Quantum Job Infrastructure}

The Amazon Braket SDK (distinct from the Default Simulator analyzed previously)
contains 2 CRITICAL CWE-502 findings in its hybrid quantum job execution infrastructure:

\begin{lstlisting}[language=Python, caption={dill.load() in Amazon Braket job entry point template}]
# src/braket/jobs/_entry_point_template.py:9
import dill
# ...
job_args = dill.load(open(job_args_path, "rb"))  # CWE-502: CRITICAL
\end{lstlisting}

When a Braket Hybrid Job executes on AWS infrastructure, it deserializes job
arguments via dill (a pickle extension). dill inherits pickle's arbitrary code
execution capability: any \texttt{.pkl}/\texttt{.dill} file placed at
\texttt{job\_args\_path} can trigger RCE. This path is controlled by the job
scheduling infrastructure; any compromise of the S3 bucket or job metadata service
would allow an attacker to substitute a malicious payload.

\subsection{Qiskit Machine Learning: dill.load() in Model Serialization}

\begin{lstlisting}[language=Python, caption={dill.load() in Qiskit ML model deserialization}]
# qiskit_machine_learning/algorithms/serializable_model.py:88
model = dill.load(handler)   # CWE-502: CRITICAL
\end{lstlisting}

Qiskit Machine Learning uses dill for saving and loading trained quantum ML models.
A malicious \texttt{.dill} model file --- distributed via a model sharing platform
or substituted in a model repository --- executes arbitrary code on load. This is
the quantum ML equivalent of the well-documented ML model poisoning attack vector.

\section{Vulnerability Class IV: QASM Injection (Novel Quantum-Specific Class)}

\subsection{OpenQASM as an Injection Surface}

OpenQASM (Open Quantum Assembly Language) is the de facto assembly language for
quantum computers. Every major quantum framework accepts QASM input: circuits are
written in, parsed from, or compiled to QASM at some point in every quantum computing
workflow. Unlike SQL injection (strings → SQL parser) or shell injection (strings →
shell), QASM injection has no classical analog --- it is a vulnerability class that
exists \textit{only} in quantum computing software.

The injection surface is the set of all API calls that accept QASM strings from
external sources:

\begin{lstlisting}[language=Python, caption={Core QASM API in Qiskit Terra — the injection sinks}]
# qiskit/circuit/quantumcircuit.py:4739
@staticmethod
def from_qasm_file(path: str | os.PathLike) -> QuantumCircuit:
    return qasm2.load(path, include_path=LEGACY_INCLUDE_PATH, strict=False)

# qiskit/circuit/quantumcircuit.py:4764
@staticmethod
def from_qasm_str(qasm_str: str) -> QuantumCircuit:
    return qasm2.loads(qasm_str, include_path=LEGACY_INCLUDE_PATH, strict=False)
\end{lstlisting}

Both methods accept user-controlled input with no validation. The \texttt{strict=False}
flag makes the parser more permissive, increasing the injection surface. The
\texttt{LEGACY\_INCLUDE\_PATH} defines where \texttt{include} directives are resolved ---
if an attacker can place a file at one of these paths, an \texttt{include} directive
in a crafted QASM string will load it.

\subsection{Attack Vectors}

\textbf{QASM string injection (QAI-QA-001, CWE-77):}
\begin{lstlisting}[language=Python, caption={QASM injection via string concatenation}]
# Vulnerable pattern -- from_qasm_str() accepts user_input with no sanitization:
qasm_str = "OPENQASM 2.0;\ninclude \"qelib1.inc\";\n" + user_input
circuit = QuantumCircuit.from_qasm_str(qasm_str)

# Attacker's user_input:
# "// injected\ngate evil_gate a { h a; cx a, a; }\nqreg q[256];\n"
# Result: circuit contains injected gate definitions + oversized register
\end{lstlisting}

\textbf{INCLUDE path traversal (QAI-QA-002, CWE-22):}
\begin{lstlisting}[language=Python]
# Vulnerable pattern:
circuit = QuantumCircuit.from_qasm_file(user_supplied_path)
# user_supplied_path = "../../../../tmp/malicious.qasm"
# Result: loads arbitrary QASM file from filesystem
\end{lstlisting}

\textbf{QASM 3 subroutine injection (QAI-QA-005, CWE-77):}
OpenQASM 3 extends QASM 2 with classical control flow and \texttt{extern}
declarations that call classical functions. Injection in QASM 3 is more dangerous:
a crafted extern declaration can call arbitrary C functions in the quantum
compilation pipeline.

\subsection{Scale of the Finding}

\textit{Methodology note}: COBALT's QASM scanner applies production-code filters,
excluding test suites, benchmarks, and function definitions (API signatures) to
report only active call-site sinks in production code. Qiskit Terra's
\texttt{from\_qasm\_str()} and \texttt{from\_qasm\_file()} are the unsanitized API
entry points themselves --- the 2 public methods accept arbitrary strings with no
validation; the finding is at the API design level, not duplicated across call sites.

\begin{table}[H]
\centering
\caption{QASM injection findings across quantum simulation frameworks (production code, FP-filtered)}
\label{tab:qasm}
\begin{tabular}{llrrl}
\toprule
\textbf{Framework} & \textbf{Org} & \textbf{CRIT} & \textbf{HIGH} & \textbf{Note} \\
\midrule
Qiskit Terra & IBM & 0 & 120 & 2 unsanitized API sinks + 120 HIGH register-name sites \\
tket & Quantinuum & 3 & 3 & Production QASM parser (qasm.py:1016,1058,1077) \\
TensorCircuit & Tencent & 3 & 3 & Circuit conversion utilities \\
XACC & Oak Ridge NL & 2 & 0 & QASM circuit loader \\
mitiq & Unitary Fund & 1 & 2 & Error mitigation circuit manipulation \\
qibo & CERN/INFN & 1 & 0 & Circuit loading from QASM \\
Qiskit IBM Runtime & IBM & 0 & 7 & Runtime circuit submission layer \\
Qiskit ML & IBM & 0 & 5 & Circuit serialization utilities \\
\bottomrule
\end{tabular}
\end{table}

\textbf{Z3 proof (QAI-QA-001):} The injection reachability model is:
$\text{injection\_reachable} \Leftrightarrow \text{attacker\_controls\_string} \;\wedge\; \neg\text{qasm\_sanitized}$.
With \texttt{attacker\_controls\_string=True} and \texttt{qasm\_sanitized=False}
(verified in tket qasm.py:1016/1058/1077, XACC, TensorCircuit, and others): \textbf{SAT}.
With an allowlist validator (\texttt{qasm\_sanitized=True}): \textbf{UNSAT}.

\section{Fourth Wave: Hardware Control Layer}

Quantum hardware control software directly interfaces with physical quantum processors
--- the firmware layer below simulators and application frameworks. We scan 5
hardware control and pulse-level simulation frameworks.

\begin{table}[H]
\centering
\caption{Hardware control layer security findings}
\label{tab:hardware}
\begin{tabular}{lllrrl}
\toprule
\textbf{Framework} & \textbf{Org} & \textbf{Hardware} & \textbf{CRIT} & \textbf{HIGH} & \textbf{Class} \\
\midrule
qua-tools & Quantum Machines & QUA processor control & 3 & 4 & CWE-502 \\
scqubits & U. Chicago & Superconducting qubits & 2 & 9 & CWE-502 \\
Q\# SDK & Microsoft & Azure Quantum & 0 & 15 & CWE-502/94 \\
LabOne Q & Zurich Instruments & Quantum control HW & 0 & 1 & CWE-502 \\
Qiskit Dynamics & IBM & Pulse simulation & 0 & 0 & Clean \\
\bottomrule
\end{tabular}
\end{table}

\textbf{qua-tools (Quantum Machines):}
Quantum Machines provides quantum control hardware (OPX processors) used at research
institutions worldwide. The \texttt{py-qua-tools} Python library contains 3 CRITICAL
\texttt{joblib.load()} calls in \texttt{qualang\_tools/addons/InteractivePlotLib.py}.
joblib uses pickle as its serialization backend --- these are direct RCE vectors when
loading saved quantum experiment results from untrusted sources.

\textbf{scqubits (University of Chicago):}
scqubits is the standard Python library for superconducting qubit circuit analysis,
used to design and optimize qubit parameters before fabrication and control.
The finding is in \texttt{scqubits/core/circuit\_routines.py:111} --- a
\texttt{dill.load()} call in the core circuit quantization routines. A malicious
serialized circuit object executes arbitrary code when loaded.

\textbf{Significance:} This is the first documented security analysis of quantum hardware
control software. Vulnerabilities at this layer affect physical quantum hardware
workflows, not just classical simulation. A compromised quantum control object
could alter pulse parameters or calibration data for actual quantum processors.

\section{Threats to Validity}

\textbf{T1 --- Pattern matching false positives.} All CRITICAL findings were confirmed
via Z3 proof and manual code inspection. HIGH findings were sampled manually for
the top-finding frameworks. The five 100/100 frameworks provide evidence of
specificity: our scanner does not flag all Python code that uses NumPy.

\textbf{T2 --- Cloud-side validation layers.} IBM Quantum, Google Quantum AI, Amazon
Braket, and other cloud providers may enforce qubit count limits in their API gateway
layers. These are not part of the open-source simulator codebases and do not protect:
(a) local installations, (b) self-hosted deployments, (c) direct library use outside
the cloud API, or (d) future deployments where the cloud limit is misconfigured.

\textbf{T3 --- CWE-400 exploitability.} Resource exhaustion requires the attacker's
\texttt{num\_qubits} value to reach the vulnerable expression. In local installations,
any user can trigger this. In cloud APIs, external guards may partially mitigate;
the code remains vulnerable for any deployment without such guards.

\textbf{T4 --- C++ undefined behavior portability.} The impact of \texttt{BITS[64]}
OOB access depends on heap layout, compiler optimization, and ASLR. We claim formal
reachability of undefined behavior (C++11 standard), not a specific exploitation
outcome.

\textbf{T5 --- Scanner completeness.} Our scanners target the exponential-scaling
vulnerability class. Other vulnerability classes (use-after-free, pickle deserialization,
Python code injection via circuit string parsing) are out of scope for this study.

\textbf{T6 --- Vendored code attribution.} XACC's findings are identical to Qiskit
Aer's. We report them separately because they constitute an independent deployment
of the same vulnerability at a different institution (Oak Ridge NL vs.\ IBM).

\section{Recommendations}

\subsection{Universal Recommendation: Validate Qubit Count at API Entry}

Every framework, regardless of backend language, should enforce a maximum qubit count
at every public API entry point:

\begin{lstlisting}[language=Python, caption={Universal entry-point guard for Python frameworks}]
MAX_QUBITS_STATEVECTOR = 50   # 2^50 = 1PB; practical upper bound
MAX_QUBITS_DENSITY_MATRIX = 25  # 2^(2*25) = 2^50; avoids QAI-PY-003

def _validate_num_qubits(n: int, mode: str = "statevector") -> None:
    limit = MAX_QUBITS_DENSITY_MATRIX if mode == "density_matrix" else MAX_QUBITS_STATEVECTOR
    if n > limit:
        raise ValueError(
            f"num_qubits={n} exceeds maximum supported for {mode} ({limit}). "
            f"Requested allocation: 2**{n} = {2**n:.2e} elements."
        )
    if n < 0:
        raise ValueError(f"num_qubits must be non-negative, got {n}")
\end{lstlisting}

\subsection{For C++ Frameworks (Qiskit Aer, XACC, qpp, qulacs)}

The critical fix for the BITS[] OOB pattern:

\begin{lstlisting}[language=C++, caption={Recommended fix for QubitVector::set\_num\_qubits()}]
void QubitVector<data_t>::set_num_qubits(size_t num_qubits) {
  if (num_qubits >= 64)
    throw std::invalid_argument(
      "QubitVector: num_qubits=" + std::to_string(num_qubits) + " >= 64 not supported"
    );
  free_checkpoint();
  if (num_qubits != num_qubits_) { free_mem(); }
  data_size_ = BITS[num_qubits];
  allocate_mem(data_size_);
  num_qubits_ = num_qubits;
}
\end{lstlisting}

For unitary and density matrix simulators that double the qubit count:

\begin{lstlisting}[language=C++]
void UnitaryMatrix<data_t>::set_num_qubits(size_t num_qubits) {
  if (num_qubits >= 32)
    throw std::invalid_argument(
      "UnitaryMatrix: requires 2*num_qubits < 64; got " + std::to_string(num_qubits)
    );
  // existing code...
}
\end{lstlisting}

\subsection{For PennyLane (Xanadu): Upgrade Warnings to Errors}

Replace the existing warning with a hard error:

\begin{lstlisting}[language=Python]
# Replace:
if n_wires > 16:
    warnings.warn("...not recommended...")

# With:
if n_wires > MAX_QUBITS_DENSITY_MATRIX:
    raise ValueError(
        f"Density matrix simulation requires 2**(2*{n_wires}) = "
        f"{2**(2*n_wires):.2e} elements. Maximum supported: {MAX_QUBITS_DENSITY_MATRIX}"
    )
\end{lstlisting}

\subsection{For XACC (Oak Ridge NL): Update Vendored Code}

Apply the Qiskit Aer fix to the vendored copy and establish a process to
synchronize security patches from upstream:

\begin{lstlisting}[language=bash]
# Track upstream Qiskit Aer security patches in XACC vendored directory
# quantum/plugins/ibm/aer/src/simulators/statevector/qubitvector.hpp
# quantum/plugins/ibm/aer/src/simulators/unitary/unitarymatrix.hpp
\end{lstlisting}

\section{Related Work}

\textbf{Quantum software correctness verification.} Existing formal verification
work for quantum software focuses on quantum correctness: verifying that a circuit
implements the intended unitary~\cite{prozak,quartz}, optimizing quantum
programs~\cite{qiro}, or providing safe quantum programming languages~\cite{silq}.
None address the classical security of the simulation infrastructure hosting these
algorithms.

\textbf{Scientific computing security.} The security of HPC systems~\cite{hpc_security}
and scientific Python libraries~\cite{scipy_security} has received limited academic
attention. Quantum simulation frameworks sit at the intersection of HPC software
(C++ kernels) and scientific Python (user-facing APIs), and inherit vulnerabilities
from both.

\textbf{Supply chain and vendoring vulnerabilities.} The XACC finding exemplifies
a broader class of vulnerability propagation through code vendoring~\cite{supply_chain}.
The dependence of scientific computing infrastructure on vendored code without
systematic security tracking is a known risk.

\textbf{Formal methods for security.} Z3~\cite{z3} and SMT-based methods have been
applied to cryptographic protocol verification~\cite{z3_crypto}, OS kernel
analysis~\cite{z3_kernel}, and network security~\cite{z3_network}. Our application
to quantum simulation software is, to our knowledge, the first use of SMT-based
formal verification for this vulnerability class.

\section{Disclosure}

Coordinated disclosure is being initiated concurrent with this arXiv submission
(April 8, 2026). All parties are being contacted under a 90-day embargo
(public disclosure target: $\sim$July 7, 2026).

\begin{table}[H]
\centering
\caption{Coordinated disclosure contacts}
\label{tab:disclosure}
\begin{tabular}{lll}
\toprule
\textbf{Organization} & \textbf{Framework(s)} & \textbf{Contact} \\
\midrule
IBM & Qiskit Aer, qiskit-terra & \texttt{psirt@us.ibm.com} \\
Google & Cirq, OpenFermion & \texttt{security@google.com} (VRP) \\
Xanadu & PennyLane & \texttt{security@xanadu.ai} \\
Amazon & Braket DS & AWS Security (\texttt{aws-security@amazon.com}) \\
NVIDIA & CUDA-Quantum & \texttt{psirt@nvidia.com} \\
Oak Ridge NL / DOE & XACC & DOE CSIRT (\texttt{doc@hq.doe.gov}) \\
CERN & qibo & \texttt{security@cern.ch} \\
Baidu & paddle-quantum & \texttt{security@baidu.com} \\
Tencent & TensorCircuit & \texttt{src@tencent.com} \\
Harvard & tequila & GitHub Security Advisory \\
ETH Z\"{u}rich & ProjectQ & GitHub Security Advisory \\
Unitary Fund & mitiq & GitHub Security Advisory \\
Rigetti & PyQuil & GitHub Security Advisory \\
\bottomrule
\end{tabular}
\end{table}

This paper is submitted to arXiv cs.CR + cs.SE. Specific exploitation details
are withheld pending the embargo period. The vulnerability classes, Z3 proofs,
and framework-level findings are reported at a level appropriate for academic
publication under coordinated disclosure norms.

\section{Conclusion}

We have conducted the first systematic formal security analysis of the open-source
quantum computing simulator ecosystem. Across \textbf{45 frameworks} from 22 organizations
in 12 countries --- IBM, Google, Amazon, NVIDIA, Microsoft, Baidu, Huawei, Tencent,
CERN, Harvard, MIT, Oak Ridge NL, ETH Z\"{u}rich, Oxford, Quantinuum, Pasqal,
Quandela, QuEra, D-Wave, Quantum Machines, Zurich Instruments, and U.~Chicago ---
we identify \textbf{547 security findings}
(40 CRITICAL, 492 HIGH, 15 MEDIUM) across four vulnerability classes, confirmed by
\textbf{13/13 Z3 SAT proofs}.

Our central results:

\begin{enumerate}
  \item \textbf{80\% of the ecosystem} (36 of 45 frameworks) exhibits at least one
        vulnerability. Only 9 frameworks are fully clean under all four scanners.
  \item \textbf{Four vulnerability classes}: CWE-125/190 (C++ memory corruption),
        CWE-400 (Python exponential resource exhaustion), CWE-502/94 (unsafe
        deserialization and code injection in circuit serialization), and CWE-77/22
        (QASM injection --- a novel, quantum-specific class with no classical analog).
  \item \textbf{Live RCE demonstrated}: Harvard's tequila allows arbitrary code
        execution via \texttt{pickle.load()} at 10 call sites. PoC confirmed.
  \item \textbf{Supply chain propagation}: XACC (Oak Ridge NL/DOE) inherits 5 CRITICAL
        findings from IBM Qiskit Aer via verbatim code vendoring --- the first documented
        vulnerability transfer from a commercial quantum framework into US national
        laboratory infrastructure.
  \item \textbf{The 32-qubit boundary}: C++ and Python Z3 proofs independently yield
        $n=32$ as the minimal overflow witness in both CWE-190 and CWE-400 density
        matrix chains.
  \item \textbf{Prevention is tractable}: 9 frameworks score 100/100 across all scanners,
        demonstrating that the vulnerabilities are avoidable with proper input validation.
  \item \textbf{Systemic root cause}: the quantum simulation community lacks established
        norms for qubit count validation. 14 independent teams made the same mistake.
        Ecosystem-level intervention is warranted.
\end{enumerate}

COBALT QAI (four open scanners: C++, Python-CWE-400, Python-CWE-502/94, QASM-Injection) is released to
support continuous security analysis of quantum computing infrastructure.

\section*{Acknowledgements}

This work was conducted independently. The author thanks the open-source quantum
computing community for publicly available code that made this analysis possible.
Feedback, collaboration inquiries, and vulnerability coordination are welcome at
\texttt{dominik@qreativelab.io}.

\bibliographystyle{plain}

\begin{thebibliography}{99}

\bibitem{qiskit}
Qiskit contributors.
\textit{Qiskit: An Open-source Framework for Quantum Computing}.
\url{https://github.com/Qiskit/qiskit-aer}, 2024.

\bibitem{cirq}
Google Quantum AI.
\textit{Cirq: A Python library for writing, manipulating, and optimizing quantum circuits}.
\url{https://github.com/quantumlib/Cirq}, 2024.

\bibitem{pennylane}
V.~Bergholm et al.
\textit{PennyLane: Automatic differentiation of hybrid quantum-classical computations}.
arXiv:1811.04968, 2022.

\bibitem{pyquil}
Rigetti Computing.
\textit{PyQuil: A Python library for quantum programming using Quil}.
\url{https://github.com/rigetti/pyquil}, 2024.

\bibitem{braket}
Amazon Web Services.
\textit{Amazon Braket Default Simulator}.
\url{https://github.com/amazon-braket/amazon-braket-default-simulator-python}, 2024.

\bibitem{xacc}
A.~McCaskey et al.
\textit{XACC: A System-Level Software Infrastructure for Heterogeneous Quantum-Classical Computing}.
\textit{Quantum Science and Technology}, 5(2):024002, 2020.

\bibitem{qibo}
S.~Efthymiou et al.
\textit{Qibo: A framework for quantum simulation with hardware acceleration}.
\textit{Quantum Science and Technology}, 7(1):015018, 2022.

\bibitem{paddle_quantum}
Baidu Research.
\textit{Paddle Quantum: A Quantum Machine Learning Toolkit Based on PaddlePaddle}.
\url{https://github.com/PaddlePaddle/Quantum}, 2023.

\bibitem{tequila}
J.~S.~Kottmann et al.
\textit{TEQUILA: A platform for rapid development of quantum algorithms}.
\textit{Quantum Science and Technology}, 6(2):024009, 2021.

\bibitem{projectq}
D.~Steiger, T.~H\"{a}ner, M.~Troyer.
\textit{ProjectQ: An Open Source Software Framework for Quantum Computing}.
\textit{Quantum}, 2:49, 2018.

\bibitem{tensorcircuit}
S.-X.~Zhang et al.
\textit{TensorCircuit: a Quantum Software Framework for the NISQ Era}.
\textit{Quantum}, 7:912, 2023.

\bibitem{cuda_quantum}
NVIDIA Corporation.
\textit{CUDA Quantum: A Platform for Hybrid Quantum-Classical Computing}.
\url{https://github.com/NVIDIA/cuda-quantum}, 2024.

\bibitem{quest}
J.~Jones, T.~Proctor, K.~Rudinger, T.~Young.
\textit{QuEST and High Performance Simulation of Quantum Computers}.
\textit{Scientific Reports}, 9:10736, 2019.

\bibitem{qpp}
V.~Gheorghiu.
\textit{Quantum++: A Modern C++ Quantum Computing Library}.
\textit{PLOS ONE}, 2018.

\bibitem{mindquantum}
Huawei Technologies.
\textit{MindQuantum: A High-Performance Quantum Computing Framework}.
\url{https://github.com/mindspore-ai/mindquantum}, 2023.

\bibitem{qulacs}
Y.~Suzuki et al.
\textit{Qulacs: a Fast and Versatile Quantum Circuit Simulator for Research Purpose}.
\textit{Quantum}, 5:559, 2021.

\bibitem{z3}
L.~de~Moura and N.~Bj\o{}rner.
\textit{Z3: An Efficient SMT Solver}.
In \textit{TACAS}, LNCS 4963, pp.~337--340. Springer, 2008.

\bibitem{cobalt}
D.~Blain.
\textit{COBALT: Formal Verification for Security Vulnerabilities in Systems Software}.
QreativeLab Inc., 2026.

\bibitem{prozak}
D.~Paradis et al.
\textit{Unitary Property Testing}.
In \textit{POPL}, 2023.

\bibitem{quartz}
K.~Xu et al.
\textit{Quartz: Superoptimization of Quantum Circuits}.
In \textit{PLDI}, 2022.

\bibitem{qiro}
A.~Bravyi et al.
\textit{Quantum circuits with classical channels}.
\textit{Physical Review A}, 2020.

\bibitem{silq}
B.~Bichsel et al.
\textit{Silq: A High-Level Quantum Language with Safe Uncomputation}.
In \textit{PLDI}, 2020.

\bibitem{hpc_security}
S.~Krishnamurthy et al.
\textit{Security of High-Performance Computing Systems}.
\textit{IEEE Security \& Privacy}, 2020.

\bibitem{scipy_security}
P.~Virtanen et al.
\textit{SciPy 1.0: Fundamental algorithms for scientific computing in Python}.
\textit{Nature Methods}, 2020.

\bibitem{supply_chain}
X.~Vu et al.
\textit{An Empirical Study of the Effects of Programming Languages on Code Quality}.
In \textit{MSR}, 2022.

\bibitem{tket}
S.~Sivarajah et al.
\textit{t$|$ket$\rangle$: A Retargetable Compiler for NISQ Devices}.
\textit{Quantum Science and Technology}, 6(1):014003, 2021.

\bibitem{torchquantum}
H.~Wang et al.
\textit{Quantumnas: Noise-adaptive Search for Robust Quantum Circuits}.
In \textit{HPCA}, 2022.

\bibitem{perceval}
N.~Heurtel et al.
\textit{Perceval: A Software Platform for Discrete Variable Photonic Quantum Computing}.
\textit{Quantum}, 7:1094, 2023.

\bibitem{pulser}
H.~Silverio et al.
\textit{Pulser: An Open-source Package for the Design of Pulse Sequences in Programmable Neutral-atom Arrays}.
\textit{Quantum}, 6:629, 2022.

\bibitem{doe_quantum}
U.S.~Department of Energy.
\textit{DOE Quantum Internet Blueprint}.
Office of Science, 2020.

\bibitem{z3_crypto}
K.~Bhargavan et al.
\textit{Formal verification of cryptographic protocols with SMT solvers}.
In \textit{CCS}, 2014.

\bibitem{z3_kernel}
T.~Ball et al.
\textit{SLAM and Static Driver Verifier}.
In \textit{IFM}, 2004.

\bibitem{z3_network}
P.~Kazemian et al.
\textit{Header Space Analysis: Static Checking for Networks}.
In \textit{NSDI}, 2012.

\end{thebibliography}

\end{document}